\documentclass[a4paper,amssymb,eqsecnum,showpacs,epsfig,useAMS]{jpconf}
\usepackage{graphicx}
\usepackage{bm}
\usepackage{amsmath}
\usepackage{hyperref}

\def \lleq {\lower0.9ex\hbox{ $\buildrel < \over \sim$} ~}
\def \ggeq {\lower0.9ex\hbox{ $\buildrel > \over \sim$} ~}

\def \beq  {\begin{equation}}
\def \eeq  {\end{equation}}
\def \ber  {\begin{eqnarray}}
\def \eer  {\end{eqnarray}}

\begin{document}
\newcommand{\newc}{\newcommand}
\newcommand{\ben}{\begin{eqnarray}}
\newcommand{\een}{\end{eqnarray}}
\newc{\be}{\begin{equation}}
\newc{\ee}{\end{equation}}
\newc{\ba}{\begin{eqnarray}}
\newc{\ea}{\end{eqnarray}}
\newc{\bea}{\begin{eqnarray*}}
\newc{\eea}{\end{eqnarray*}}
\newc{\D}{\partial}
\newc{\ie}{{\it i.e.} }
\newc{\eg}{{\it e.g.} }
\newc{\etc}{{\it etc.} }
\newcommand{\nn}{\nonumber}
\newc{\ra}{\rightarrow}
\newc{\lra}{\leftrightarrow}

\title{Genetic algorithms and the analysis of SnIa data}

\author{Savvas Nesseris}

\address{The Niels Bohr International Academy and DISCOVERY Center, The Niels Bohr Institute, Blegdamsvej 17,
DK-2100}

\ead{nesseris@nbi.dk}

\begin{abstract}
The Genetic Algorithm is a heuristic that can be used to produce
model independent solutions to an optimization problem, thus making it ideal
for use in cosmology and more specifically in the analysis of type Ia supernovae
data. In this work we use the Genetic Algorithms (GA) in order
to derive a null test on the spatially flat cosmological constant
model $\Lambda$CDM. This is done in two steps: first, we apply the
GA to the Constitution SNIa data in order to acquire a model
independent reconstruction of the expansion history of the
Universe $H(z)$ and second, we use the reconstructed $H(z)$ in
conjunction with the Om statistic, which is constant only for the
$\Lambda$CDM model, to derive our constraints. We find that while
$\Lambda$CDM is consistent with the data at the $2\sigma$ level,
some deviations from $\Lambda$CDM model at low redshifts can be accommodated.
\end{abstract}

\section{Introduction}
In the previous decade it was discovered that the Universe is
undergoing an accelerated expansion \cite{Riess:2004nr},
\cite{Spergel:2006hy}, \cite{Readhead:2004gy}. This acceleration is usually
attributed either to a cosmic fluid with negative pressure dubbed
Dark Energy or to an IR modification of gravity. In order to
identify the properties of Dark Energy or the
structure of the IR modification of gravity it is necessary to
know to a high precision the rate of the expansion of the
Universe, parameterized as $H\equiv\frac{\dot{a}}{a}$ where
$a=\frac{1}{1+z}$ is the scale factor and $z$ is the redshift of
the cosmological probe, as it measured by the observations.

The behavior of the expansion of the Universe can be identified by
studying two functions, the Equation of State (EoS) $w(z)\equiv
\frac{P}{\rho}$ which can be rewritten as \be
w(z)\,=\,-1\,+\frac{1}{3}(1+z)\frac{d\ln (\delta H(z)^2)}{d
z}\,,{\label{wzh1}} \ee where $\delta
H(z)^2=H(z)^2/H_0^2-\Omega_{\rm 0m} (1+z)^3$ accounts for all
terms in the Friedmann equation not related to matter and the
deceleration parameter $q(z)\equiv -\frac{\ddot{a}}{a \dot{a}^2}$
which can be rewritten as \be q(z)\,=\,-1\,+(1+z) \frac{d\ln
(H(z))}{d z}\,,{\label{qzh1}} \ee Obviously, the cosmological
constant ($w(z)=-1$) corresponds to a constant dark energy
density, while in general $w(z)$ can be time dependent. Also, an
important parameter is the value of the deceleration parameter
today, ie $q(z=0)\equiv q_0$, which for the cosmological constant
model in GR it is $q_0=-1+3 \Omega_{\rm 0m}/2$.

However, despite all the recent progress the origin of the
accelerated expansion of the universe still remains unknown with
many possibilities still remaining open, see for example
\cite{Perivolaropoulos:2006ce}. The simplest choice that agrees
well with the data is a positive cosmological constant which has
to be small enough to have started dominating the universe at late
times. As it was demonstrated by the Seven-Year WMAP data
\cite{Komatsu:2010fb}, the cosmological constant remains the best
candidate and has the advantage of having only one free parameter
related to the properties of the Dark Energy. Nonetheless, this
model fails to explain why the cosmological constant is so small
that it can only dominate the universe at late times, a problem
known as the {\it coincidence problem} and there are a few
cosmological observations which differ from its predictions
\cite{Perivolaropoulos:2008ud},\cite{Perivolaropoulos:2008yc}.

A very important complication in the investigation of the behavior
of dark energy occurs due to the bias introduced by the
parameterizations used. At the moment, there is a multitude of
available phenomenological ans\"atze for the dark energy equation
of state parameter $w$ or dark energy density, each with its own
merits and limitations (see \cite{Sahni:2006pa} and references
therein). The interpretation of the SNIa data has been shown to
depend greatly on the type of parametrization used to perform a
data fit \cite{Sahni:2006pa}, \cite{arman}.
Choosing a priori a model for dark energy can thus adversely affect the
validity of the fitting method and lead to compromised or misleading results.

The need to counteract this problem paved the way for the
consideration of a complementary set of non-parametric
reconstruction techniques \cite{Daly:2003iy}, \cite{Wang:2003gz},
\cite{Saini:2003pa}, \cite{Wang:2009sn}, \cite{Clarkson:2010bm}  and model independent approaches
(\cite{statefinder},\cite{Alam03}, \cite{arman}, \cite{arman2},
\cite{wang_teg05},\cite{SC10}. These try to minimize
the ambiguity due to a possibly biased assumption for $w$ by
fitting the original datasets without using any parameters related
to some specific model. The result of these methods can then be
interpreted in the context of a dark energy model of choice.
Non-parametric reconstructions can thus corroborate parametric
methods and provide more credibility. However, they too suffer
from a different set of problems, mainly the need to resort to
differentiation of noisy data, which can itself introduce great
errors.

In this paper, we present a method that can be used as a model
independent approach in testing the standard cosmological model.
This is done by using the Genetic Algorithms (GA) technique, first
used in the analysis of SNIa data in \cite{Bogdanos:2009ib}.
The GAs represent a method for non-parametric reconstruction of
the dark energy equation of state parameter $w$, based on the
notions of genetic algorithms and grammatical evolution. GAs are
more useful and efficient than usual techniques when
\begin{itemize}
    \item The parameter space is very large, too complex or not enough
    understood, as is the case with dark energy.
    \item Domain knowledge is scarce or expert knowledge is difficult to encode to narrow the search space.
    \item Traditional search methods give poor results or completely fail.
\end{itemize} Naturally, therefore, they have been used with success in many fields where
one of the above situations is encountered, like the computational
science, engineering and economics. Recently, they have also been
applied to study high energy physics
\cite{Becks:1994mm},\cite{Allanach:2004my}, \cite{Rojo:2004iq}
gravitational wave detection \cite{Crowder:2006wh} and
gravitational lensing \cite{Brewer:2005ww}. Since the nature of
Dark Energy still remains a mystery, this makes it for us an ideal
candidate to use the GAs as a means to analyze the SNIa data and
extract model independent constraints on the behavior of the Dark
Energy.

This talk was delivered at the 14th conference in the series ``Recent 
Developments in Gravity" (NEB-14) and is a brief overview of the use 
of the Genetic Algorithms to a model independent reconstruction
of the expansion history of the Universe. In Section 2 a brief description 
of the Om statistic is given while in Section 3 we provide an overview 
of the general methodology of the GA paradigm and finally, we present 
our results in Section 4.

\section{The Om statistic}
The recently introduced $Om$ diagnostic \cite{om} (also look
at Ref.~\cite{litmus}) enables us to distinguish $\Lambda$CDM from
other dark energy models without directly involving the cosmic
EOS. The $Om$ diagnostic is defined as:
\begin{equation}
Om(x) \equiv \frac{h^2(x)-1}{x^3-1},\hspace{6 mm} x=1+z~,
\hspace{6 mm} h(x) = H(x)/H_0~. \label{eq:om}
\end{equation}
For dark energy with a constant equation of state $w=const$,

\begin{equation}
h^2(x) = \Omega_{0m}x^3 + (1-\Omega_{0m})x^\alpha,\hspace{6 mm}
~~\alpha = 3(1+w) \label{eq:hubble}
\end{equation}

\noindent(we assume that the universe is spatially flat for
simplicity). Consequently,
\begin{equation}
Om(x) = \Omega_{0m} + (1-\Omega_{0m})\frac{x^\alpha - 1}{x^3-1}~,
\end{equation}

\begin{figure}
\centering
\includegraphics[height=.4\textheight]{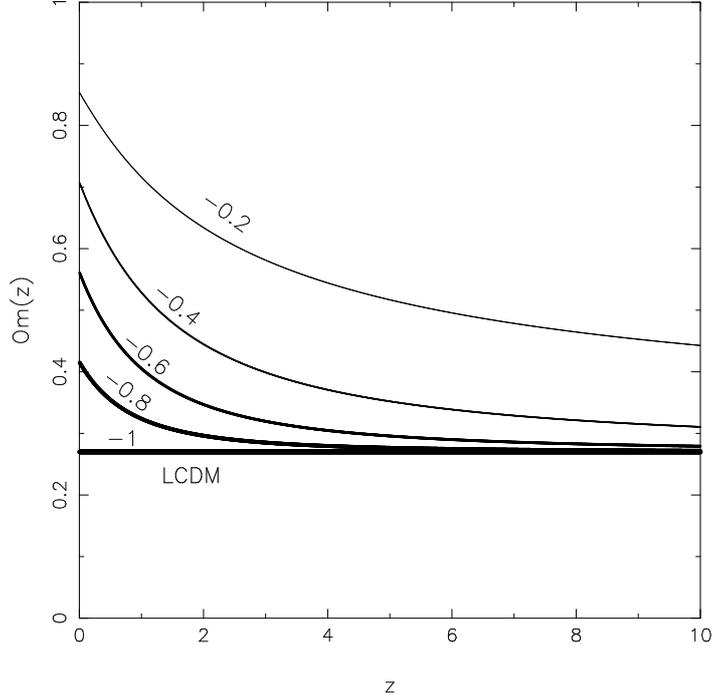}
\caption{The $Om$ diagnostic is shown as a function of redshift
for dark energy models with $\Omega_{0m}=0.27$ and $w= -1, -0.8,
-0.6, -0.4, -0.2$ (bottom to top). For Phantom models (not shown)
$Om$ would have the opposite curvature.\label{fig:om}}
\end{figure}

\noindent from where we find $Om(x) = \Omega_{0m}$ in
$\Lambda$CDM, whereas $Om(x) > \Omega_{0m}$ in quintessence
$(\alpha > 0)$ while $Om(x) < \Omega_{0m}$ in phantom $(\alpha <
0)$. We therefore conclude that: $Om(x) - \Omega_{0m} = 0$ if dark
energy is a cosmological constant. Note that since
$\Omega_\Lambda+\Omega_{0m} \simeq 1$
 in $\Lambda$CDM, this model contains
SCDM $(\Omega_{0m}=1, \Omega_\Lambda=0)$ as an important limiting
case. Consequently, the $Om$ diagnostic cannot distinguish between
large and small values of the cosmological constant unless the
value of the matter density is independently known. In other
words, the $Om$ diagnostic provides us with a {\em null test} of
the cosmological constant. This is a simple consequence of the
fact that $h^2(x)$ plotted against $x^3$ results in a straight
line for $\Lambda$CDM, whose slope is given by $\Omega_{0m}$. For
other dark energy models the line describing $Om(x)$ is curved,
since the equality

\begin{equation}
\frac{d[h^2]}{d[x^3]} = constant~,
\label{eq:slope}
\end{equation}
(which always holds for $\Lambda$CDM for any $x=1+z$) is satisfied
in quintessence/phantom type models only at redshifts
significantly greater than unity, when the effects of dark energy
on the expansion rate can safely be ignored. As a result the
efficiency of the $Om$ diagnostic improves at low redshifts ($z <
2$) precisely where there is likely to be an abundance of
cosmological data in the coming years!

For a constant EoS: $1+w \simeq [Om(z) -
\Omega_{0m}](1-\Omega_{0m})^{-1}~$ at $z \ll 1$, consequently a
larger $Om(z)$ is indicative of a larger $w$; while at high $z$,
$Om(z) \to \Omega_{0m}$, as shown in Fig.~\ref{fig:om}.

On the other hand, for quintessence as well as phantom the line
describing $Om(x)$ is curved, which helps distinguish these models
from $\Lambda$CDM even if the value of the matter density is not
accurately known -- see Fig.~\ref{fig:om}.

In practice, the construction of $Om$ requires a knowledge of the
Hubble parameter, $h(z)$, which can be determined using a number
of model independent approaches
\cite{statefinder},\cite{Alam03}, \cite{arman}, \cite{arman2},
\cite{wang_teg05},\cite{SC10}.

\section{Genetic algorithms}
\subsection{Overview}
GAs were introduced as a computational analogy of adaptive
systems. They are modeled loosely on the principles of the
evolution via natural selection, employing a population of
individuals that undergo selection in the presence of
variation-inducing operators such as  mutation and crossover. The
encoding of the chromosomes is called the genome or genotype and
is composed, in loose correspondence to actual DNA genomes, by a
series of representative ``genes''. Depending on the problem, the
genome can be a series (vector) of binary numbers, decimal
integers, machine precision integers or reals or more complex data
structures.

In order to evaluate the individual chromosomes a fitness function
is used and reproductive success usually varies with fitness. The
fitness function is a map between the gene sequence of the
chromosomes (genotype) and a number of attributes (phenotype),
directly related to the properties of a wanted solution. Often the
fitness function is used to determine a ``distance'' of a
candidate solution from the true one. Various distance measures
can be used for this purpose (Euclidean distance, Manhattan,
Mahalanobis etc.).

The algorithm begins with an initial population of candidate
solutions, which is usually randomly generated. Although GA's are
relatively insensitive to initial conditions, i.e. the population
we start from is not very significant, but using some prescription
for producing this seed generation can affect the speed of
convergence. In each successive step, the fitness functions for
the chromosomes of the population are evaluated and a number of
genetic operators (mutation and crossover) are applied to produce
the next generation. This process continues until some termination
criteria is reached, e.g. obtain a solution with fitness greater
than some predefined threshold or reach a maximum number of
generations. The later is imposed as a condition to ensure that
the algorithm terminates even if we cannot get the desired level
of fitness.

The various steps of the algorithm can be summarized as follows:
\begin{enumerate}
    \item Randomly generate an initial population $M(0)$
    \item Compute and save the fitness for each individual m in the current population
    $M(t)$.
    \item Define selection probabilities $p(m)$ for each individual $m$ in $M(t)$ so that $p(m)$ is proportional to
    the fitness.
    \item Generate $M(t+1)$ by probabilistically selecting individuals
    from $M(t)$ to produce offspring via genetic operators (crossover and mutation).
    \item Repeat step 2 until a satisfying solution is obtained, or a maximum number of generations reached.
\end{enumerate}

We should stress that the initial population $M(0)$ will only
depend on the choice of the grammar and therefore it can only
affect on how fast the algorithm will converge to the minimum.
Using the wrong grammar may result in the GA being trapped in a
local minimum. Also, two important parameters that affect the
transition from the population $M(t)$ to $M(t+1)$ are the
selection rate and the mutation rate. The selection rate is
typically of the order of $10\%$ and it affects how many of the
individuals, after they are ranked with respect to their fitness,
will be allowed to produce offspring. The mutation rate is usually
of the order of $5\%$ and it expresses the probability that an
arbitrary part of the genetic sequence will be changed from its
previous state. This is done in order to maintain the genetic
diversity from one generation of a population to the next.

The paradigm of GAs described above is usually the one applied to
solving most of the problems presented to GAs. Though it might not
find the best solution, more often than not, it would come up with
a partially optimal solution. A more detailed overview and its
application in cosmology can be found in \cite{Bogdanos:2009ib}.

The difference of the GA over the the standard analysis of the
data, ie having an a priori defined model with some free
parameters, is that the obtained values of the best-fit parameters
are model-dependent and in general models with more parameters
tend to give better fits to the data. This is where the GA
approach starts to depart from the ordinary parametric method. Our
goal is to minimize a function, not using a candidate model
function for the distance modulus and varying parameters, but
through a stochastic process based on a GA evolution. This way, no
prior knowledge of a dark energy model is needed to obtain a
solution and our result will be completely parameter-free.

In simpler terms, the GA method does not require an a priori
assumption for a DE model, but uses the data themselves to find
this model. Also, it is parameter free as the end result does not
have any free parameters that can be changed in order to fit the
data. So, in this sense this method has far less bias any of the
standard methods for the reconstruction of the expansion history
of the Universe. This is the main reason for the use of the GAs
in this paper.

\subsection{General Methodology}
We first outline the course of action we follow to apply the GA
paradigm in the case of SNIa data. For the application of GA and
grammatical evolution (GE) on the dataset, we use a modified
version of the GDF \cite{Tsoulos} tool
~\footnote{\href{http://cpc.cs.qub.ac.uk/summaries/ADXC}{http://cpc.cs.qub.ac.uk/summaries/ADXC}},
which uses GE as a method to fit datasets of arbitrary size and
dimensionality. This program uses the tournament selection method
for crossover. GDF requires a set of input data (train set), an
(optional) test sample and a grammar to be used for the generation
of functional expressions. The output is an expression for the
function which best fits the train set data.

Since the SNIa datapoints are given in terms of the distance
modulus $\mu_{obs}(z_i)$, the fitness function for the GA we have
chosen is equal to $-\chi^{2}_{SNIa}$, where \be \chi^2_{SNIa}=
\sum_{i=1}^N \frac{(\mu_{obs}(z_i) - \mu_{GA}(z_i))^2}{\sigma_{\mu
\; i}^2 } \label{chi2}\,, \ee where for the Constitution set
$N=397$ and $\sigma_{\mu \; i}^2$ are the errors due to flux
uncertainties, intrinsic dispersion of SNIa absolute magnitude and
peculiar velocity dispersion and $\mu_{GA}(z_i)$ is the {\it
reduced distance modulus} obtained for each chromosome of the
population by the GA.

An advantage of our method is the fact that neither the
$\chi^2_{SNIa}$ nor the $\mu_{GA}(z)$ depend on any parameters.
Also, as it can be seen from eq. (\ref{chi2}), our method does not
make an assumption of flatness or not in order to fit the data.
Flatness (or non-flatness) only comes into play when one tries to
find the luminosity distance $d_L(z)$ and the underlying dark
energy model given the best-fit form of $\mu_{GA}(z)$. So, in this
aspect the result produced by the GA is independent of the
assumption of flatness as well.

The GA evaluates (\ref{chi2}) in each evolutionary step for every
chromosome of the population. The one with the best fitness will
consequently have the smallest $\chi^2_{SNIa}$ and will be the
best candidate solution in its generation. Of all the steps in the
execution of the algorithm, the evaluation of the fitness is the
most expensive. GDF is appropriately modified to use (\ref{chi2})
as the basis of its fitness calculation.

After the execution of the GA, we obtain an expression
$\mu_{GA}(z)$ for the reduced distance modulus as the solution of
best fitness and the corresponding  $\chi^2_{SNIa}$. Using this we
can, through differentiation, obtain other functions or parameters
of interest, such as the Om statistic or the deceleration
parameter $q(z)$. For example, for a flat Universe the Hubble
parameter will be given by \be
1/H(z)=\frac{d}{dz}\left(\frac{10^{\frac{\mu_{GA}(z)}{5}}}{1+z}\right)
\ee and then the deceleration parameter can be found by
Eq.~(\ref{qzh1}).

A possible complication of this method is that it gives no direct
way to estimate the errors for the derived parameters. One cannot
expect to get an estimate by just running the algorithm many times
and obtaining slightly different parameters. The GA usually tends
to converge at the same solution for a given dataset, unless we
change significantly the population size or the number of
generations. A way to circumvent this problem is to use a
bootstrap Monte Carlo simulation to produce synthetic datasets and
rerun the algorithm on them. We can thus obtain a statistical
sample of parameter values which will allow us to estimate the
error.

We sketch the procedure we will follow \cite{press92}:
\begin{enumerate}
\item The GA is applied on the original SNIa dataset with the
chosen execution parameters and a solution for $\bar \mu_{GA}(z)$
is obtained. \item Generate a number of synthetic datasets by
drawing each time the same number of data points with replacement
from the original set. \item The GA is rerun for the synthetic
datasets. \item A new set of values for the Om statistic and the
deceleration parameter $q(z)$ is generated. \item The $95\%$ error
of the desired parameter can be found by taking the 2.5 and the
97.5 percentiles of the bootstrap distribution
\cite{Efron:1982xx}.
\end{enumerate}
Using the above steps, we can obtain error estimates for any
desired parameter.

\subsection{An example}

In this Section we will briefly describe a simple
example\footnote{This is only a schematic description of how the
GA works and this is done solely for the sake of explaining the
basic mechanisms of the GA.} of how he GA determines the best-fit.
We will avoid describing the technicalities, like the binary
representation of the solutions, and instead we will concentrate
on how the generations evolve, how many and which individuals are
chosen for the next generation etc.

As we mentioned earlier, the choice of the grammar is very
important, however in order to keep our example as simple as
possible we will assume that our grammar includes only basic
functions like polynomials $x$, $x^2$ etc, the trigonometric
functions $sin(x)$, $cos(x)$, the exponential $e^x$ and the
logarithm $ln(x)$.

The first step in the GA is setting up a random initial population
$M(0)$ which can be any simple combination of these functions, eg
$\mu_{GA,1}(z)=ln(z)$, $\mu_{GA,2}(z)=-1+z+z^2$ and
$\mu_{GA,3}(z)=sin(z)$. The number of number of candidate
solutions (chromosomes) in the genetic population is usually a few
hundreds and later on we will use the value of 500.

Next, the algorithm measures the fitness of each solution by
calculating their $\chi^2$, ie for our simple example
$\chi^2_1=80579.9$, $\chi^2_2=292767.0$ and $\chi^2_3=412928.0$.
The selection per se is done by implementing the ``Tournament
selection" method which involves running several ``tournaments",
sorting the population with respect to the fitness each individual
and after that a fixed percentage, adjusted by the selection rate
(see the previous subsection), of the population is chosen. As we
mentioned, the selection rate is of the order of $10\%$ of the
total population, but lets assume that the two out of the three
candidate solutions ($\mu_{GA,1}(z)$ and $\mu_{GA,2}(z)$) are
chosen for the sake of simplicity.

The reproduction of these two solutions will be done by the
crossover and mutation operations. The crossover will randomly
combine parts of the ``parent'' solutions, for example in one such
realization this may be schematically  shown as\ba
\mu_{GA,1}(z)\oplus\mu_{GA,2}(z)&\rightarrow &
(\bar\mu_{GA,1}(z),\bar\mu_{GA,2}(z),\bar\mu_{GA,3}(z)) \nn
\\&=&\nn\left(ln(z^2),-1 + ln(z^2),-1 + ln(z)\right) \ea

After this is done, the GA will proceed to implement the mutation
operation. The probability for mutation is adjusted by the
mutation rate, which as we mentioned is typically of the order of
$5\%$. In our example this may be a change in the power of some
term or the change in the value of a coefficient. For example, for
the candidate solution $\bar\mu_{GA,3}(z)$ this can be
schematically shown as $\bar\mu_{GA,3}(z)=-1 + ln(z)\rightarrow -1
+ ln(z^3)$, where the power of the $z$ term was mutated from $1$
to $3$.

Finally, at the end of the first round we have the three candidate
solutions \ba
M(1)&=&(\bar\mu_{GA,1}(z),\bar\mu_{GA,2}(z),~~\bar\mu_{GA,3}(z))= \nn \\
&&\left(ln(z^2),-1 + ln(z^2),-1 + ln(z^3)\right) \nn\ea At the
beginning of the next round the fitness of each individual will
again be determined, which for our population will be
$(\chi^2_1,\chi^2_2,\chi^2_3) =(3909.2,18435,113327)$, and the
selection and the other operations will proceed as before. It is
easily seen that even after one generation the $\chi^2$ of the
candidate solutions has decreased dramatically and as we will see
in the next section, it only takes a bit more than 100 generations
to reach an acceptable $\chi^2$ and about 500 to 1000 generations
to start converging on the minimum.

After a predetermined number of generations, usually on the order
of $10^3$, has been reached or some other criteria have been met,
the GA will finish, with the best candidate solution of the last
generation being the best-fit to the data.

\section{Results}

We will now examine a number of different configurations,
highlighted in Table \ref{table1}. In order to keep our analysis
as simple as possible we will use a Basic Grammar that includes
polynomials like $x$, $x^2$ etc, the trigonometric functions
$sin(x)$, $cos(x)$, the exponential $e^x$ and the logarithm
$ln(x)$ and a second grammar that includes the Basic Grammar and
the Legendre polynomials. For the analysis we will use the
Constitution SNIa dataset of \cite{Hicken:2009dk}
consisting of 397 SNIa. The steps followed for the usual
minimization of (\ref{chi2}) in terms of its parameters are
described in detail in \cite{Nesseris:2004wj}, \cite{Nesseris:2005ur},\cite{Lazkoz:2005sp},
\cite{Nesseris:2006er}.

\begin{figure}[h]
\begin{minipage}{19pc}
\includegraphics[width=1\textwidth]{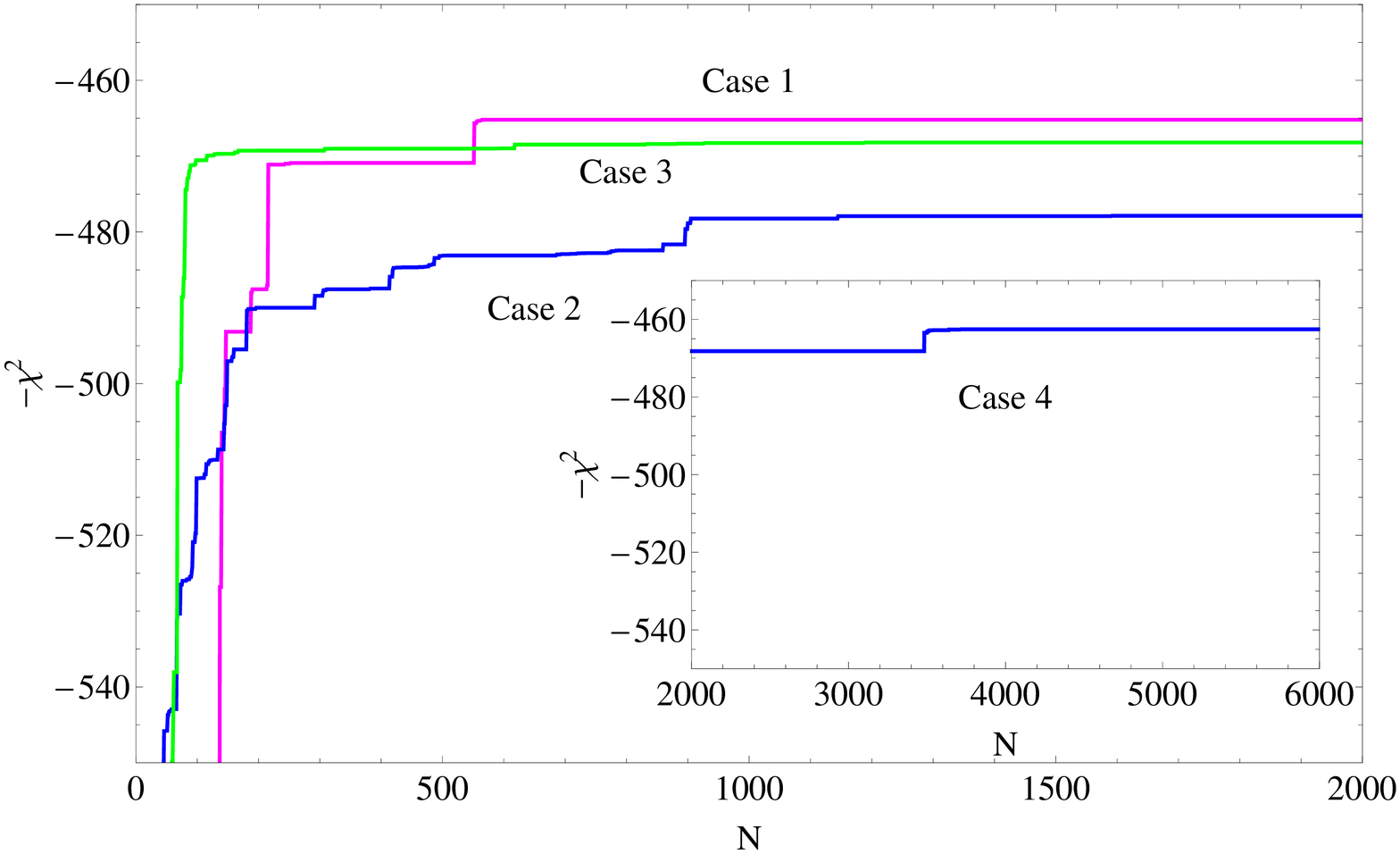}
\caption{The fitness (equal to $-\chi^2$) as a function of the number of generations. The magenta line corresponds to case 1, the
green line to case 2 and the blue line to case 3. The inset graphic shows case 4.\label{fig2}}
\end{minipage}\hspace{1pc}%
\begin{minipage}{19pc}
\includegraphics[width=1\textwidth]{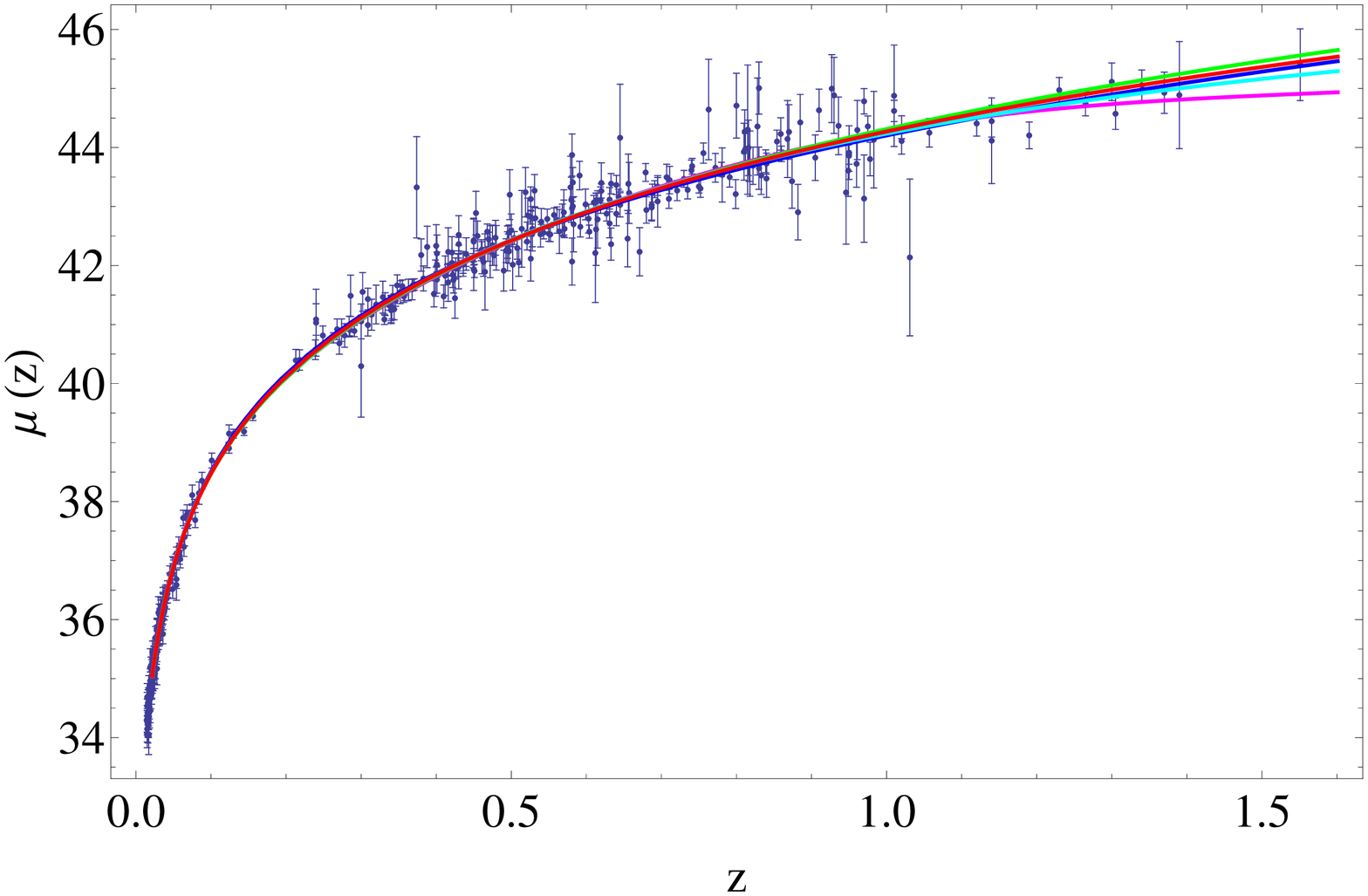}
\caption{The SNIa distance modulus against the redshift. The red line represents the best-fit $\Lambda$CDM model with $\Omega_{m}=0.289$ while the magenta, blue, green and cyan (best-fit) lines correspond to cases 1-4 respectively.\label{fig3}}
\end{minipage}
\end{figure}

In Fig.~\ref{fig2} we show the fitness (equal to $-\chi^2$) as a
function of the number of generations. Clearly, all configurations
seem to have to have converged very early or to be very close to
converging to their minimum. An exception is Case 3 which even
after 3000 generations still has not converged and we had to run
the simulation for another 3000 generations to check its
convergence. Since we want to compare the configurations with
equal standards, ie number of generations etc, we will treat this
case (Case 4) separately.

As it can be seen from Table \ref{table1} and Fig.~\ref{fig3},
most configurations agree well with the data and actually have a
minimum $\chi^2$ comparable or even better with respect to
$\Lambda$CDM. An exception to this is Case 2, where the best fit
differs from $\Lambda$CDM by a $\Delta \chi^2=12$ thus providing a
much poorer fit. The reason for this is the fact that by increasing
the grammar we effectively increase the landscape in which the
algorithm has to search and therefore, we are making
the fit poorer as the GA will take much much longer to converge.

\vspace{0pt}
\begin{table}
\begin{center}
\caption{The different cases considered in the analysis. The Basic
Grammar includes polynomials, trigonometric functions,
exponentials and logarithms. For $\Lambda$CDM the best-fit
corresponds to $\Omega_{m}=0.289$. \label{table1}}
\begin{tabular}{ccc}
\hline
\hline\\
\vspace{6pt}  \textbf{Case 1} & Basic grammar & $\chi^2_{min}=465.19$\\
\vspace{6pt}  \textbf{Case 2} & Basic grammar + Legendre polyn.  \hspace{6pt}& $\chi^2_{min}=477.87$\\
\vspace{6pt}  \textbf{Case 3} & Only polyn. in grammar (3000 gen.) & $\chi^2_{min}=468.19$\\
\vspace{6pt}  \textbf{Case 4} & Only polyn. in grammar (6000 gen.) & $\chi^2_{min}=462.56$\\
\vspace{6pt}  \textbf{$\Lambda$CDM} & $H(z)^2=H_0^2(\Omega_{m} (1+z)^3+1-\Omega_{m} )$ & $\chi^2_{min}=465.51$\\
 \hline \hline
\end{tabular}
\end{center}
\end{table}

\begin{figure}[h]
\begin{minipage}{19pc}
\includegraphics[width=1\textwidth]{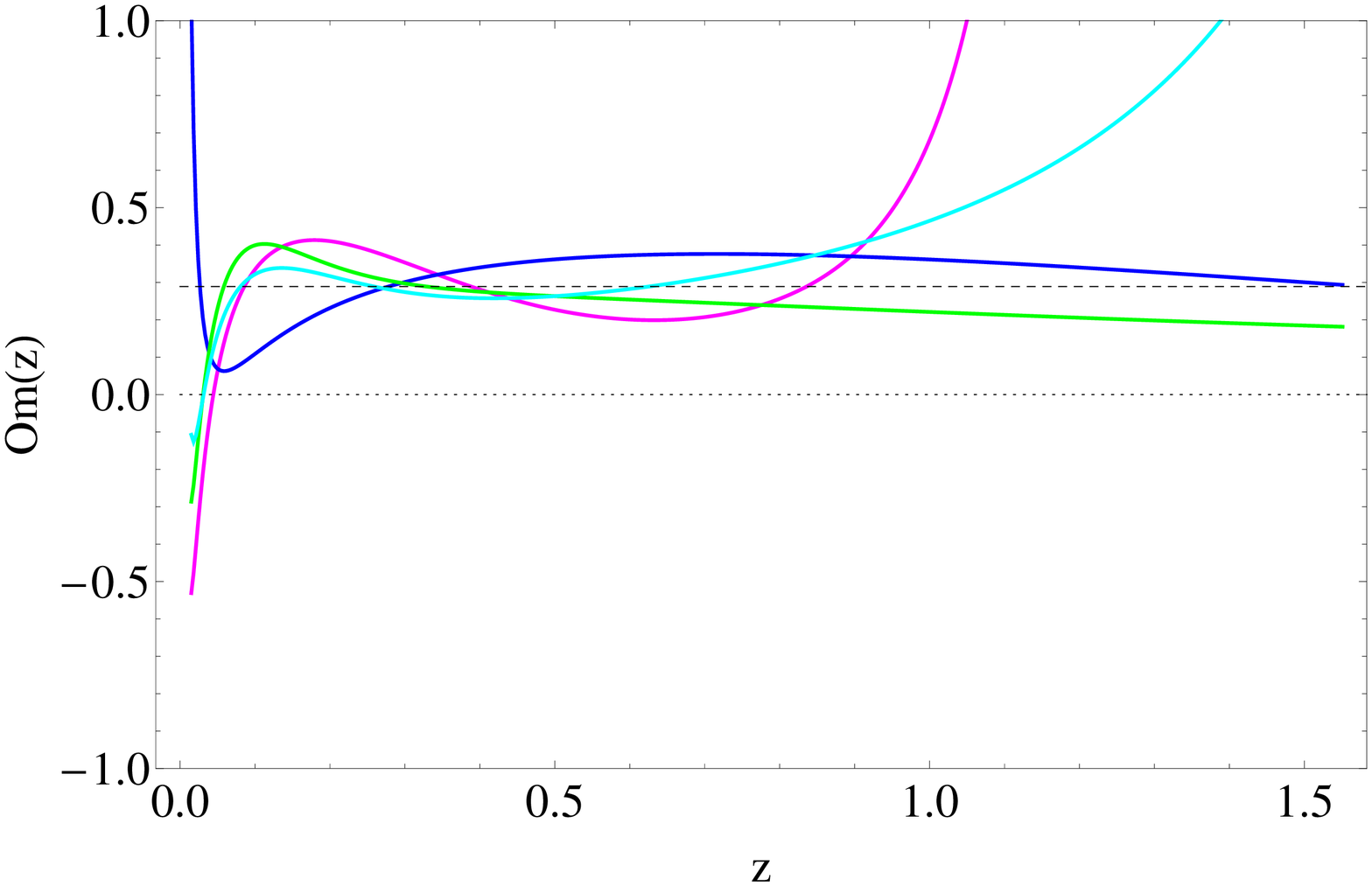}
\caption{The Om statistic against the redshift. The black dashed
line represents the $\Lambda$CDM value of $\Omega_{m}=0.289$ while
the magenta, blue, green and cyan (best-fit) lines  correspond to cases 1-4
respectively.\label{fig4}}
\end{minipage}\hspace{1pc}%
\begin{minipage}{19pc}
\includegraphics[width=1\textwidth]{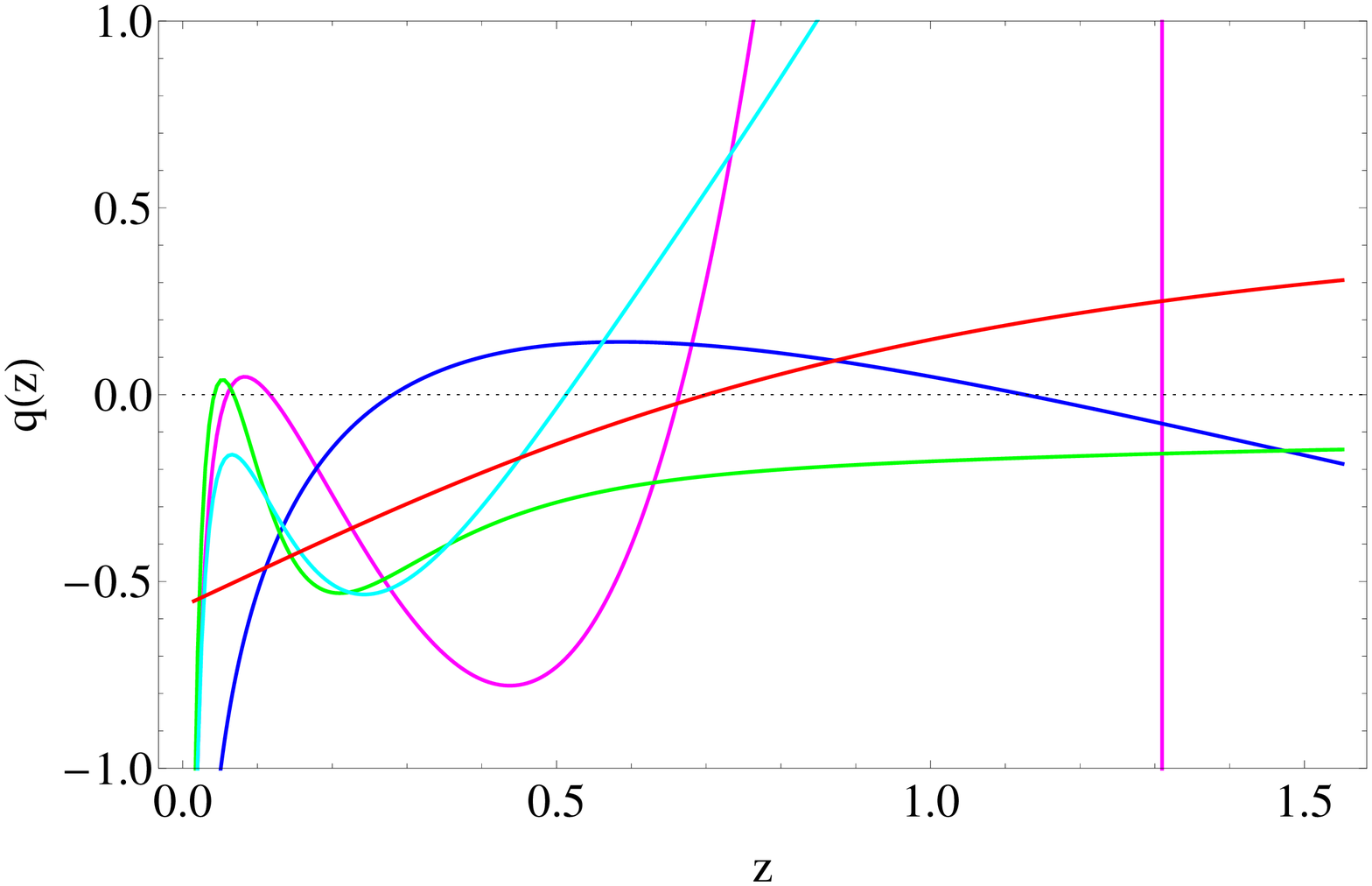}
\caption{The deceleration parameter $q$ against the redshift. The
red line represents the best-fit $\Lambda$CDM model with
$\Omega_{m}=0.289$ while the magenta, blue, green and cyan (best-fit) lines
correspond to cases 1-4 respectively.\label{fig5}}
\end{minipage}
\end{figure}

In Figs.~\ref{fig4} and ~\ref{fig5} we show the corresponding
results of each case for the Om statistic and the deceleration
parameter $q(z)$ (our best-fit function corresponds to the cyan line).
Due to the presence of trigonometric functions into the best fit
and the fact that the Om statistic is derived from the Hubble
parameter by differentiation, Case 1 exhibits strong oscillations.
This is also apparent in the corresponding plot for the deceleration
parameter for Case 2 (magenta line, Fig.~\ref{fig5}).

As it can be seen in Fig.~\ref{fig:om}, a constant $w$ (but different from $w=-1$)
forces the Om statistic to be a monotonic function that is located
completely either above or below the constant case ($Om=0.289$).
However, none of the curves in Fig.~\ref{fig4} are  monotonic functions and
all of them cross the constant Om case several times. Also, all cases remain reasonably
close to the $\Lambda$CDM value of $\Omega_{m}=0.289$ for intermediate
redshifts but deviate from that value at high or low redshifts. Therefore,
this could be the effect of a time-dependent equation of state $w(z)$
instead of a constant $w$. Comparing with Fig.~\ref{fig:om}, it is
possible that this behaviour is due to an equation of state
$w(z)$ more negative than -1.

This is also supported from Fig.~\ref{fig4} and Fig.~\ref{fig5} as at low redshifts all
cases predict a much different forms of acceleration than the best
fit $\Lambda$CDM.  In Fig.~\ref{fig4} and Fig.~\ref{fig5} our best
fit results show an overall slowing down of the acceleration at
the low redshifts (which is in concordance with results presented
in \cite{arman} with a very rapid increase of acceleration at the
very low redshifts. These are very similar with the results
presented in Fig.2 of Ref.~ \cite{SC10} where a smoothing method for the
reconstruction of the expansion history of the universe is being
used. It is obvious that when two very different model independent
approaches reach to a similar result, we must be on the right
track and very close to the best possible result. However, theoretical
interpretation of these results is beyond the scope of this paper.
In fact there might be very different mechanisms that result to a
similar expansion history of the universe as we reconstructed in this
paper (e.g see \cite{CKT}, \cite{Alam03}, \cite{Kelly09}, \cite{Sullivan09} ).

\begin{figure}[h]
\begin{minipage}{19pc}
\includegraphics[width=1\textwidth]{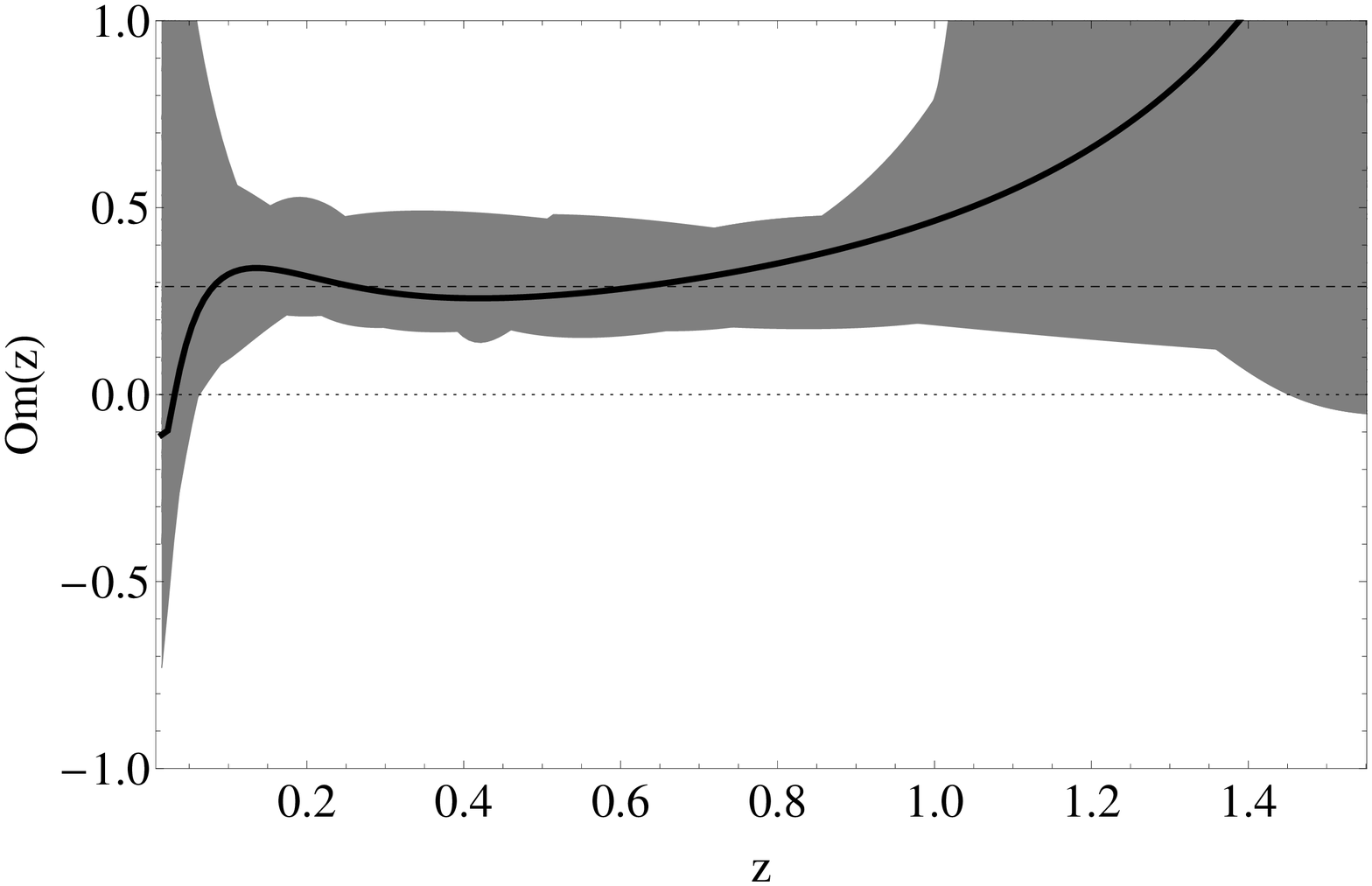}
\caption{The Om statistic as a function of $z$. The black
line corresponds to case 4 (the best-fit) and the gray-shaded
area to the $2\sigma$ error. \label{fig6}}
\end{minipage}\hspace{1pc}%
\begin{minipage}{19pc}
\vspace{-0.35cm}
\includegraphics[width=1\textwidth]{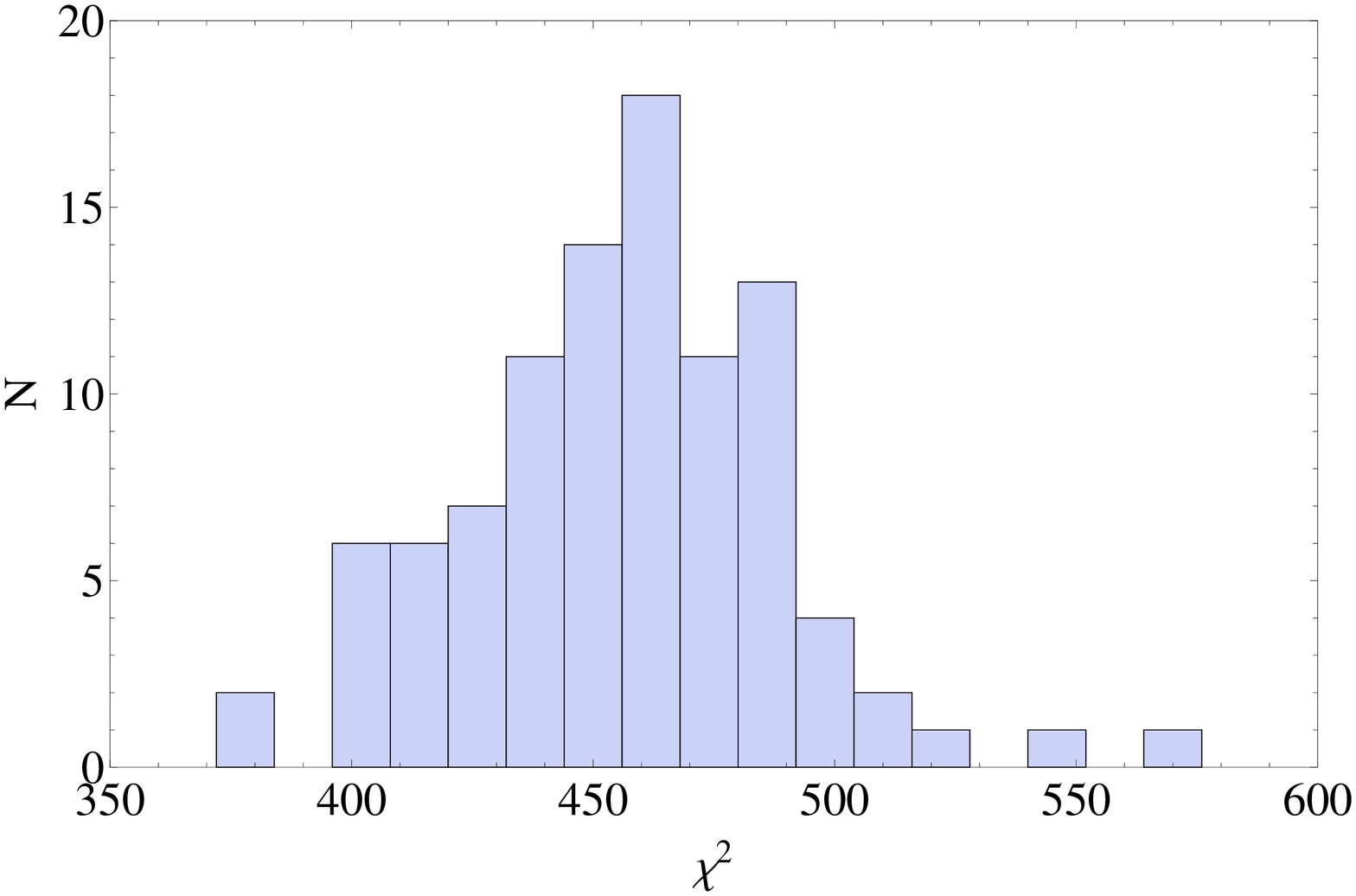}
\caption{Histogram of the bootstrap distribution found from the
monte-carlo simulation that was used to create the error regions
of Fig.~\ref{fig6}. \label{fig7}}
\end{minipage}
\end{figure}

In Fig.~\ref{fig6} we show the Om statistic as a function of the
redshift $z$. The black line corresponds to the best fit of case
4, while the gray-shaded area to the $2\sigma$ error region. The
error region was calculated by implementing a bootstrap
monte-carlo simulation as discussed in the previous section. Note
that the overall slowing down of the acceleration at the low
redshifts cannot be clearly seen from this figure as the errors
are quite symmetrical around the best fit $\Lambda$CDM value.
However, as we mentioned earlier this is clearly seen by the
behavior of the deceleration parameter in Fig.~\ref{fig5}.

As it can be seen in Fig.~\ref{fig6}, $\Lambda$CDM remains
consistent with the data at the $2\sigma$ level but at the same
time, some deviations from the standard $\Lambda$CDM model can be
accommodated. Due to the large errorbars of the current datasets, many
dark energy models are still consistent with them but we expect
this to change in the near future with the advent of high quality SnIa
observations.

In Fig.~\ref{fig7} we show a histogram of the bootstrap
distribution found from the monte-carlo simulation that was used
to create the error regions. Clearly, the distribution is quite
symmetrical and centered, so we are confident that our error
region in Fig.~\ref{fig6} was correctly reconstructed. Finally, we also
tested the GA on a mock $\Lambda$CDM dataset and we found that it can
reconstruct the luminosity distance curve to within $0.1\%$,
so we are confident that is does not overfit the data, but instead it
can successfully detect the underlying DE model.

\section{Conclusions}
We used the Om statistic and the GAs in order to derive a null
test on the cosmological constant model $\Lambda$CDM. Our interest
in the GAs stems from the fact that they represent a method for
non-parametric reconstruction of the expansion history of the
universe, based on the notions of genetic algorithms and
grammatical evolution. These kinds of algorithms are more useful
and efficient than usual techniques especially when the problem
under examination is not well understood, as is the case with dark
energy. Since the nature of dark energy still remains a mystery,
this makes it for us an ideal candidate to use the GAs as a means
to analyze the SNIa data and extract model independent constraints
on the behavior of the Dark Energy. On the other hand, the $Om$
diagnostic (Sahni et al. 2008) enables us to distinguish $\Lambda$CDM from
other dark energy models without directly involving the cosmic
EoS.

Our methodology is completely model independent and it can be
summarized in two steps: first, we applied the GA to the
Constitution SNIa data in order to acquire a model independent
reconstruction of the expansion history of the Universe $H(z)$.
After we reconstructed $H(z)$, the second step was to use it in
conjunction with the Om statistic and derive our null test.

Our main result was that $\Lambda$CDM remains consistent with the
data at the $2\sigma$ level, see Fig.~\ref{fig6}, but at the same
time, some deviations from the standard $\Lambda$CDM model can be
accommodated and this is also in accordance with the data especially
at low redshifts (see the behavior of the best fit in Fig.~\ref{fig6}).

However, it should be mentioned that the slowing down at low redshifts
for the Constitution set mentioned earlier can also be seen with
the standard analysis with the CPL ansatz (see \cite{Sanchez:2009ka})
but only for the Constitution set. For
the other SNIa datasets the reverse trend (speeding up the
acceleration at low z) was observed. For this reason and due to
the fact that the current data still have quite large errors we
believe that the full potential of our method, while it is very
promising, will only be realized in the near future when more high
quality SNIa data will become available. When this happens, the
error region at low $z$ in Fig.~\ref{fig6} will be small enough
for the Om statistic to completely discriminate between
$\Lambda$CDM and the various Dark Energy models.

\section*{Acknowledgements}
S.N. is supported by the Niels Bohr International Academy, the
Danish Research Council under FNU Grant No. 272-08-0285 and the DISCOVERY center.
The current work was presented on the 14th Conference on \textit{Recent Developments in
Gravity} (NEB14) and is based on Ref.~\cite{Nesseris:2010ep}.


\section*{References}
\begin{thebibliography}{99}
\bibitem{Riess:2004nr}
Riess A.~G.~ et al.  [Supernova Search Team Collaboration],
2004, ApJ, \textbf{607}, 665.

\bibitem{Spergel:2006hy}
Spergel D.~N.~ et al.  [WMAP Collaboration],
 2007, ApJ.Sup.,  \textbf{170}, 377,
  [arXiv:astro-ph/0603449].

\bibitem{Readhead:2004gy}
Readhead A.~C.~S.~ et al.,
2004, ApJ., \textbf{609}, 498,  [arXiv:
astro-ph/0402359].

\bibitem{Perivolaropoulos:2006ce}
Perivolaropoulos L.,
2006, AIP Conf.\ Proc.\  \textbf{848}, 698,
  [arXiv:astro-ph/0601014].

\bibitem{Komatsu:2010fb}
Komatsu E.~ et al.,
2010, arXiv:1001.4538.

\bibitem{Perivolaropoulos:2008ud}
Perivolaropoulos L.,
2008, arXiv:0811.4684 [astro-ph].

\bibitem{Perivolaropoulos:2008yc}
Perivolaropoulos L.,~ Shafieloo A.,
  Phys.\ Rev.\  D {\bf 79}, 123502 (2009)
  [arXiv:0811.2802 [astro-ph]].

\bibitem{Sahni:2006pa}
Sahni V.,~ Starobinsky A.~,
 2006, Int.\ J.\ Mod.\ Phys.\  D \textbf{15}, 2105,
  [arXiv:astro-ph/0610026].

\bibitem{arman}
Shafieloo A., Alam U.,Sahni V.,~ Starobinsky A.~A.,
2006,  MNRAS,  \textbf{366}, 1081,
  [arXiv:astro-ph/0505329].

\bibitem{Daly:2003iy}
Daly R.~A.,~ Djorgovski S.~G.~,
2003, ApJ, \textbf{597}, 9,
  [arXiv:astro-ph/0305197].

\bibitem{Wang:2003gz}
Wang Y., Mukherjee P.,
 2004, ApJ., \textbf{606}, 654,
  [arXiv:astro-ph/0312192].

\bibitem{Saini:2003pa}
Saini T.~D.,
 2003, MNRAS.,  \textbf{344}, 129
  [arXiv:astro-ph/0302291].

\bibitem{Wang:2009sn}
Wang Y.,
2009,  Phys.\ Rev.\  D \textbf{80}, 123525,
  [arXiv:0910.2492].

\bibitem{Clarkson:2010bm}
Clarkson C.,~ Zunckel C.~,
2010,  arXiv:1002.5004.

\bibitem{statefinder}
Sahni V.,~ Saini T.~D.,~Starobinsky A.~A.,~ Alam U.~,
 2003, JETP Lett.\  \textbf{77}, 201
  [Pisma Zh.\ Eksp.\ Teor.\ Fiz.\  77, 249],
  [arXiv:astro-ph/0201498].

\bibitem{Alam03}
Alam U. , Sahni V. , Starobinsky A. A. ,
2003, JCAP, \textbf{0304}, 002

\bibitem{arman2}
Shafieloo A.,
2007, MNRAS,  \textbf{380}, 1573,
  [arXiv:astro-ph/0703034].

\bibitem{wang_teg05}
Wang Y., Tegmark M.,
 2005, Phys.\ Rev.\  D \textbf{71}, 103513,
  [arXiv:astro-ph/0501351].

\bibitem{SC10}
Shafieloo A.,~ Clarkson C.,
 2010,  Phys.\ Rev.\  D \textbf{81}, 083537,
  [arXiv:0911.4858 [astro-ph.CO]].

\bibitem{Bogdanos:2009ib}
Bogdanos C.,~ Nesseris S.~,
  2009, JCAP \textbf{0905}, 006,
  [arXiv:0903.2805 [astro-ph.CO]].

\bibitem{Becks:1994mm}
Becks K.~H.~,Hahn S.,~ Hemker A.~,
1994, Phys.\ Bl.\  \textbf{50}, 238;

\bibitem{Allanach:2004my}
Allanach B.~C.~,Grellscheid D.,~ Quevedo F.~,
  2004, JHEP, \textbf{0407}, 069,
  [arXiv:hep-ph/0406277].

\bibitem{Rojo:2004iq}
Rojo J.,~ Latorre J.~I.,~
2004, JHEP \textbf{0401}, 055,
  [arXiv:hep-ph/0401047];

\bibitem{Crowder:2006wh}
Crowder J.~,Cornish N.~J.,~ Reddinger L.~,
2006, Phys.\ Rev.\  D ,\textbf{73}, 063011,
  [arXiv:gr-qc/0601036].

\bibitem{Brewer:2005ww}
Brewer B.~J.~, Lewis G.~F.~,
 2005, arXiv:astro-ph/0501202.

\bibitem{om}
Sahni V.,~Shafieloo A.,~ Starobinsky A.~A.~,
  Phys.\ Rev.\  D {\bf 78}, 103502 (2008)
  [arXiv:0807.3548 [astro-ph]].

\bibitem{litmus}
Zunckel C., Clarkson C.,
2008,  Phys.\ Rev.\ Lett.\  \textbf{101}, 181301,
  [arXiv:0807.4304 [astro-ph]].

\bibitem{Tsoulos}
Tsoulos I.G.,~Gavrilis D.,~Dermatas E., 2006, Computer Physics
Communications \textbf{174}, 555-559.

\bibitem{press92}
Press W.~H.~ et. al., 1994, `Numerical Recipes', Cambridge
University Press.

\bibitem{Efron:1982xx}
Efron B.~,
1982,  Society of Industrial and Applied Mathematics, CBMS-NSF Monographs \textbf{38}.

\bibitem{Hicken:2009dk}
Hicken M.~et al.,
2009, ApJ, \textbf{700}, 1097,
  [arXiv:0901.4804 [astro-ph.CO]].

\bibitem{Nesseris:2004wj}
Nesseris S.,~Perivolaropoulos L.,
 2004, Phys.\ Rev.\ D \textbf{70}, 043531,
  [arXiv:astro-ph/0401556].

\bibitem{Nesseris:2005ur}
Nesseris S.,~ Perivolaropoulos L.,
 2005, Phys.\ Rev.\ D \textbf{72}, 123519,
  [arXiv:astro-ph/0511040].

\bibitem{Lazkoz:2005sp}
Lazkoz R.,~Nesseris S.,~ Perivolaropoulos L.~,
2005, JCAP \textbf{0511}, 010,
  [arXiv:astro-ph/0503230].

\bibitem{Nesseris:2006er}
Nesseris S.,~Perivolaropoulos L.,
2007, JCAP 0701, 018
  [arXiv:astro-ph/0610092].

\bibitem{CKT}
Csaki C. , Kaloper N. , Terning J. ,
 2002, Phys.\ Rev.\ Lett, \textbf{88}, 161302

\bibitem{Kelly09}
Kelly P. et al.,
2010, ApJ., \textbf{715}, 743 (2009)
  [arXiv:0912.0929 ].

\bibitem{Sullivan09}
Sullivan M. et al.,
  Mon.\ Not.\ Roy.\ Astron.\ Soc.\  {\bf 406}, 782 (2010)
  [arXiv:1003.5119 [astro-ph.CO]].

\bibitem{Sanchez:2009ka}
Sanchez J.~C.~B.~, Nesseris S.~, Perivolaropoulos L.,
 2009, JCAP \textbf{0911}, 029,
  [arXiv:0908.2636 [astro-ph.CO]].

\bibitem{Nesseris:2010ep}
  Nesseris S., Shafieloo A.,
  Mon.\ Not.\ Roy.\ Astron.\ Soc.\  {\bf 408}, 1879 (2010)
  [arXiv:1004.0960 [astro-ph.CO]].

\end{thebibliography}
\end{document}